\documentstyle[preprint,aps,psfig]{revtex}
\tighten
\begin{document}
\draft
\preprint{\vbox{Submitted to Physical Review B 
		          \hfill FSU-SCRI-98-40 \\
		          \null\hfill CU-NPL-1159 \\}}
\title{Perturbation Theory for Spin Ladders Using 
       Angular-Momentum Coupled Bases}	
\author{J. Piekarewicz}
\address{Supercomputer Computations Research Institute, \\
         Florida State University, Tallahassee, FL 32306}
\author{J.R. Shepard}
\address{Department of Physics, \\
         University of Colorado, Boulder, CO 80309}
\date{\today}
\maketitle
 
\begin{abstract}
We compute bulk properties of Heisenberg spin-1/2 ladders using
Rayleigh-Schr\"odinger perturbation theory in the rung and plaquette
bases. We formulate a method to extract high-order perturbative
coefficients in the bulk limit from solutions for relatively small
finite clusters. For example, a perturbative calculation for an
isotropic $2\times 12$ ladder yields an eleventh-order estimate of the
ground-state energy per site that is within 0.02\% of the
density-matrix-renormalization-group (DMRG) value.  Moreover, the
method also enables a reliable estimate of the radius of convergence
of the perturbative expansion. We find that for the rung basis the
radius of convergence is $\lambda_c\simeq 0.8$, with $\lambda$
defining the ratio between the coupling along the chain relative to
the coupling across the chain. In contrast, for the plaquette basis we
estimate a radius of convergence of $\lambda_c\simeq 1.25$. Thus, we
conclude that the plaquette basis offers the only currently available 
perturbative approach which can provide a reliable treatment of the 
physically interesting case of isotropic $(\lambda=1)$ spin ladders.  
We illustrate our methods by computing perturbative coefficients for 
the ground-state energy per site, the gap, and the one-magnon 
dispersion relation.
\end{abstract}

\narrowtext

\section{Introduction}
\label{sec:intro}

\begin{figure}
\centerline{
  \psfig{figure=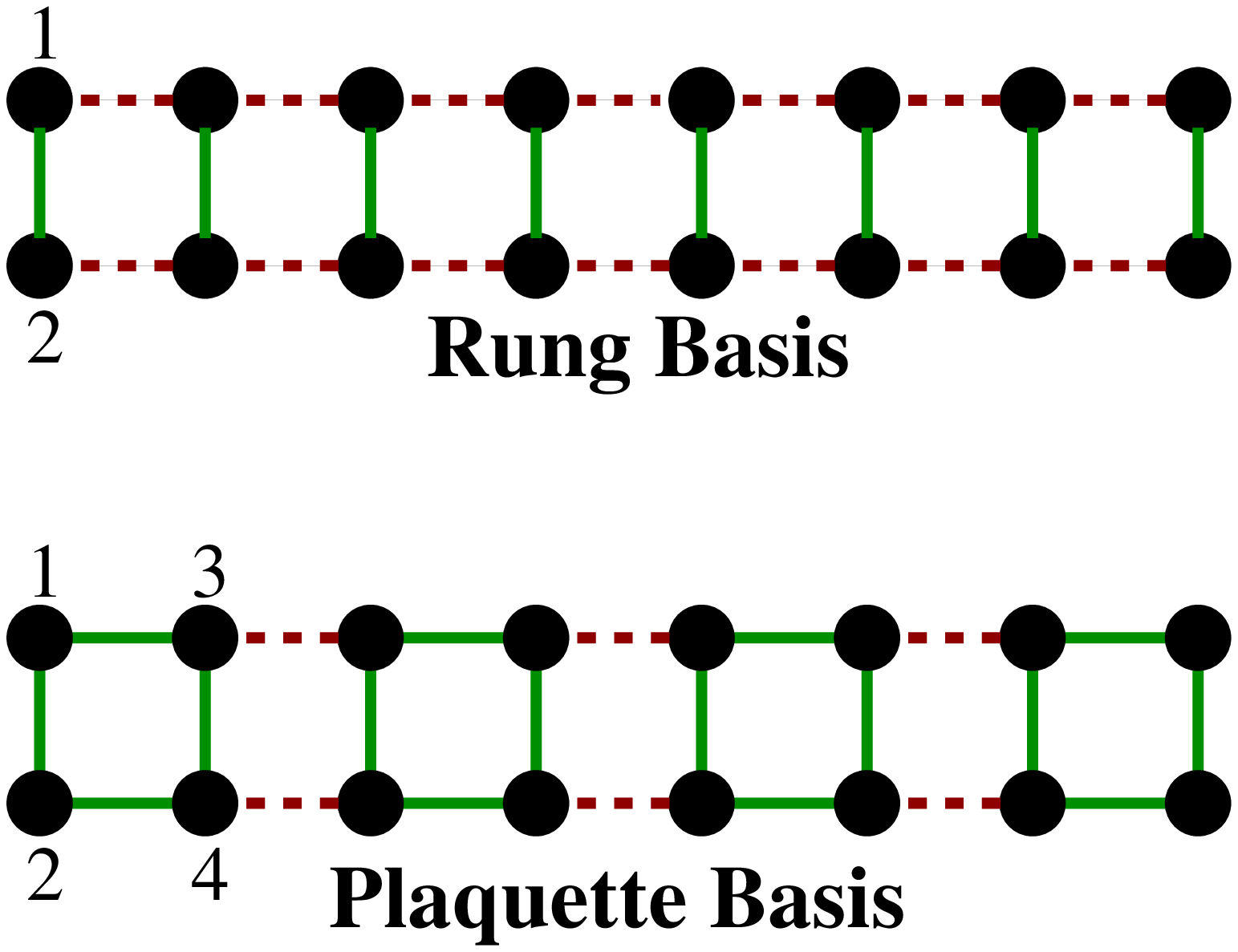,height=4in,width=5in,angle=-90}}
 \vskip 0.5in
 \caption{Schematic representation of the rung and plaquette bases.
	  The solid links represent those interactions which are diagonal 
	  in the basis; the dashed links represent the 
	  ``perturbations''.}
 \label{figzero}
\end{figure}

 Heisenberg spin ladders have been extensively studied via a number of 
powerful theoretical methods including direct 
diagonalization~\cite{barnes93}, quantum Monte Carlo~\cite{barnes93} and  
density matrix renormalization group (DMRG)~\cite{white94}.
Such investigations have been complemented
by development of simpler, often analytic, treatments whose role is to 
help elucidate the physics underlying the numerical results. These 
approaches are typically based on perturbation theory and almost invariably 
assume a ``dimer'' or ``rung'' basis (see Fig.~\ref{figzero}). Note that 
in the rung basis, the two-leg-ladder Hamiltonian becomes diagonal in the 
limit in which the coupling along the chain ($J_\parallel$) becomes 
negligible relative to the coupling across the chain ($J_\perp$). In this 
basis, the intra-chain couplings are, thus, treated perturbatively. Such 
perturbative expansions are routinely assumed to be meaningful, even in 
the region where the inter-chain and intra-chain couplings are comparable 
({\it i.e.},~$J_\parallel/J_\perp\sim 1$)~\cite{barnes93,reig94,gopa94}.

In the present work we present a perturbative treatment of two-leg
spin ladders implemented using a ``plaquette'' basis which, as we have
argued elsewhere~\cite{piek96,piek97}, is particularly well-suited to
the study of such systems for a variety of reasons. Our method
exploits the simplicity of these quasi one-dimensional systems to
straightforwardly extract perturbative information about the bulk
system from ordinary Rayleigh-Schr\"odinger perturbation theory. In so
doing we avoid the more general, and, thus necessarily more
complicated, diagrammatic methods of ``connected graph
expansions''~\cite{gelf90}. Specifically we can, with modest
computational effort, determine the bulk values of the ground state
energy per site, the spin gap, and the one magnon dispersion relation
to eleventh, fourth and third order, respectively. Moreover, by
performing similar calculations in the rung basis, we can make
reliable quantitative comparisons of the radii of convergence of the
perturbative expansion in the two bases. Indeed, we determine that the
radius of convergence in the rung basis is $\lambda_c\equiv
J_\parallel/J_\perp\simeq 0.8$ while for the plaquette basis, where,
as will be discussed below, $\lambda$ scales the inter-plaquette
coupling, we find $\lambda_c\simeq 1.25$ to $1.4$ . Clearly, the
plaquette basis is strongly preferred for the treatment of the
physically interesting\cite{dagric96} case of isotropic ({\it i.e.},
$\lambda=1$) spin ladders.

Our paper is organized as follows. In Sec.~\ref{sec:finite} we briefly
describe the rung and plaquette bases and outline how to apply
Rayleigh-Schr\"odinger perturbation theory to finite spin ladders in
these bases. In Sec.~\ref{sec:bulk} our method for extracting
perturbative information for bulk systems from the finite-ladder
results is outlined. Our numerical results appear in
Sec.~\ref{sec:results} where we also address the issue of the radii of
convergence in the rung and plaquette bases. The paper concludes with
a summary section.

\section{Perturbation Theory in Coupled Bases: Finite Systems}
\label{sec:finite}

\subsection{The Rung Basis}
\label{sec:rungbasis}
 For a two-leg ladder, the dynamics are described by the
Heisenberg Hamiltonian with nearest-neighbor antiferromagnetic
coupling:
\begin{equation}
    H=J_{\parallel}\sum_{\leftrightarrow}{\bf S}_{i}\cdot{\bf S}_{j}
     +J_{\perp}\sum_{\updownarrow}{\bf S}_{i}\cdot{\bf S}_{j} \;,
 \label{hrung} 
\end{equation}
where $J_{\perp}(J_{\parallel})$ is the strength of the
antiferromagnetic coupling along the rungs(chains) of the
ladder~\cite{barnes93}. In the strong-coupling limit
($J_{\parallel}/J_{\perp}\simeq 0$) rungs interact very weakly 
with each other and the above separation of the Hamiltonian is 
a natural one. Indeed, by fixing the large scale in the problem 
to be one ($J_{\perp}\equiv 1$) the Hamiltonian can be separated 
into the sum of $H_{0}$, which contains the intra-rung interactions, 
and a potential $V$---scaled by the small parameter
$\lambda\equiv~J_{\parallel}/J_{\perp}$, responsible for inducing
interactions between rungs. In this way, the dominant intra-rung 
Hamiltonian can be written as (see Fig.~\ref{figzero})
\begin{equation}
  H_{0}=\sum_{r=1}^{N_R}h_{0}^{(r)}\;; \quad 
  h_{0}^{(r)}={\bf S}_{1}^{(r)}\cdot{\bf S}_{2}^{(r)} \;,
 \label{hzerorung} 
\end{equation} 
while the inter-rung potential ($\lambda V$) is expressible in terms of
\begin{equation}
  V=\sum_{r=1}^{N_R}
       \left[{\bf S}_{1}^{(r)}\cdot{\bf S}_{1}^{(r+1)}
            +{\bf S}_{2}^{(r)}\cdot{\bf S}_{2}^{(r+1)}\right] \;.
 \label{rungvee} 
\end{equation} 
Note that $r$ labels a specific rung and the sum runs over all $N_{R}$ 
rungs in the ladder. 

The rung basis is particularly suitable in the strong-coupling regime,
as $H_{0}$ is diagonal in this basis. Indeed, eigenstates(eigenenergies) 
of $H_{0}$ are a direct product(sum) of single-rung states(energies)
of the form:
\begin{equation}
   |\phi_{\ell m}\rangle = 
   |(s_{1}s_{2})\,\ell m\rangle \;; \quad
   \epsilon_{\ell}={1\over 2}\ell(\ell+1)-3/4 \;.
  \label{onerung} 
\end{equation}
Here the two spins along the rung are coupled to a total angular
momentum $\ell$, which can be zero or one, and projection $m$.
What remains to be done is to evaluate matrix elements of the
potential in the rung basis. To do so, we rely on the same 
angular-momentum techniques~\cite{bieden81,brisat93} that were
used to evaluate matrix elements of the Hamiltonian in the plaquette 
basis\cite{piek96,piek97}. We obtain
\begin{eqnarray}
 &&\langle
  \ell^{\prime(1)}\ell^{\prime(2)}jm|V(1,2)|
  \ell^{(1)} \ell^{(2)} jm\rangle=\pm 3\,
  \widehat{\ell^{(1)}} \widehat{\ell^{(2)}}
  \widehat{\ell^{\prime(1)}}\widehat{\ell^{\prime(2)}}  \nonumber  \\
        &&\left\{\matrix{\ell^{(1)} & \ell^{\prime(1)}   &  1     \cr
              \ell^{\prime(2)} & \ell^{(2)}  &  j                 \cr}\right\} 
          \left\{\matrix{\ell^{(1)}  & \ell^{\prime(1)}  &  1     \cr
                             1/2     &     1/2      & 1/2    \cr}\right\} 
          \left\{\matrix{\ell^{(2)}  & \ell^{\prime(2)}  &  1     \cr
                             1/2     &     1/2      & 1/2    \cr}\right\} \;.
\end{eqnarray}
Note that we have defined $\hat{x}\equiv\sqrt{2x+1}$ and that 
the $+/-$ sign in the above expression should be adopted whenever
$\ell^{(1)}+\ell^{\prime(1)}+j$~is even/odd. Moreover, the above
expression vanishes whenever
$\ell^{(1)}+\ell^{(2)}+\ell^{\prime(1)}+\ell^{\prime(2)}$ is odd.
Aside from simple phases and numerical factors, the matrix elements 
of $V(1,2)$ depend on scalar functions (i.e., independent of $m$) 
known as Racah coefficients; here we cast the matrix elements in 
terms of the more symmetric $6\!-\!j$ symbols~\cite{bieden81,brisat93}.

\subsection{The Plaquette Basis}
\label{sec:plaquettebasis}
 For an isotropic ($J_{\parallel}=J_{\perp}\equiv1$) two-leg ladder, 
an efficient decomposition can be carried out in terms of distinct
pairs of adjacent rungs; these are the plaquettes. In such a 
decomposition, a $2\times 4$ ladder, for example, would be viewed 
as a pair of interacting plaquettes. The ladder Hamiltonian can
be written as the sum of $H_0$, which contains the intra-plaquette 
interactions, and $V$, which includes the interactions between
plaquettes. The intra-plaquette Hamiltonian $H_0$ is expressible 
as (see Fig.~\ref{figzero}) 
\begin{equation}
  H_0=\sum_{p=1}^{N_{P}}\ h^{(p)}_0 \;;\quad 
  h^{(p)}_0 = \left[{\bf S}^{(p)}_1 + {\bf S}^{(p)}_4\right]\cdot
              \left[{\bf S}^{(p)}_2 + {\bf S}^{(p)}_3\right] \;,
  \label{hzero} 
\end{equation}
where $p$ labels a specific plaquette and the sum runs over all
$N_{P}$ plaquettes in the ladder. As was shown in Ref.~\cite{piek96}, 
$H_0$ is diagonal in the plaquette basis. The single-plaquette 
wavefunctions and energies are of the form:
\begin{eqnarray}
 & |\phi_{\alpha}\rangle = 
   |(s_{1}s_{4})\ell_{14},(s_{2}s_{3})\ell_{23};jm\rangle \;, \\
 & \epsilon_{\alpha}={1\over 2}
   \bigl[j(j+1)-\ell_{14}(\ell_{14}+1)-\ell_{23}(\ell_{23}+1)\bigr] \;,
   \label{oneplaquette} 
\end{eqnarray}
where $\alpha\!\equiv\!\{\ell_{14},\ell_{23},j,m\}$ denotes a 
generic quantum label. In the plaquette basis the two {\it diagonal} 
pairs of spins are coupled to well-defined total angular momenta, 
$\ell_{14}$ and $\ell_{23}$, which can equal zero or one. These two 
link angular momenta are in turn coupled to a total plaquette angular 
momentum $j$ with projection $m$. In this scheme, the interaction $V$ 
is given by
\begin{equation}
  V=\sum_{p=1}^{N_{P}}\, \left[ 
   {\bf S}^{(p)}_3\cdot {\bf S}^{(p+1)}_1 +
   {\bf S}^{(p)}_4\cdot {\bf S}^{(p+1)}_2 \right] .
  \label{plaquettevee} 
\end{equation}
Matrix elements of $V$ in the plaquette basis are computed in terms
of five Racah coefficients, as was shown in Ref~\cite{piek96}.

\subsection{Rayleigh-Schr\"odinger Perturbation Theory}

 In this section we review some well-known results in perturbation
theory~\cite{dandf,schiff}. To formulate Rayleigh-Schr\"odinger 
perturbation theory for {\it finite ladders}, we let the Hamiltonian
$H\rightarrow H(\lambda)\equiv H_0+\lambda V$ and make a consistent 
expansion of the energy and wavefunction in powers of the ``small''
parameter $\lambda$. That is, the exact wavefunction of $H(\lambda)$
is expanded in a power series in terms of the eigenstates of $H_0$
\begin{equation}
  |\Psi_{\alpha}\rangle=
  \sum_{n=0}^\infty\lambda^n
  \sum_{\beta}A^{(n)}_{\alpha\beta}|\Phi_{\beta}\rangle \;,
  \label{pertwfn} 
\end{equation}
while the energy is expanded as
\begin{equation}
  E_{\alpha}=
  \sum_{n=0}^\infty \lambda^n\ E_{\alpha}^{(n)}.
 \label{pertenergy} 
\end{equation}
Note that $A_{\alpha\alpha}^{(n)}\!=\!\delta_{n 0}$,
$A_{\alpha\beta}^{(0)}\!=\!\delta_{\alpha\beta}$, and
$E^{(0)}_{\alpha}\!\equiv\!\epsilon_{\alpha}$ is an eigenvalue
of the unperturbed Hamiltonian $H_0$. It is straightforward to
derive a recursion relation for the above coefficients which is
well-suited for implementation on a computer to arbitrary order
in perturbation theory. We obtain
\begin{mathletters}
 \begin{eqnarray}
     E_{\alpha}^{(n)}&=&\sum_{\beta}
     \langle\Phi_{\alpha}|V|\Phi_{\beta}\rangle
     A_{\alpha\beta}^{(n-1)} \,, \\
     A_{\alpha\beta}^{(n)}&=& 
    {1 \over \epsilon_{\alpha}-\epsilon_{\beta}}
    \left[ \sum_{\sigma} A_{\alpha\sigma}^{(n-1)}
    \langle\Phi_{\sigma}|V|\Phi_{\beta}\rangle -
    \sum_{m=1}^{n-1}E_{\alpha}^{(n-m)}A_{\alpha\beta}^{(m)}\right] \;.
 \end{eqnarray}
 \label{pertn}
\end{mathletters}
These formulas are very simple to apply to the ground state of a
$2\times N_{R}$ ladder consisting of $N_{P}=N_{R}/2$ plaquettes.
The unperturbed state is given by
\begin{equation}
  |\Phi_{J=0}\rangle = 
  |\phi_0\rangle \otimes |\phi_0\rangle \ldots |\phi_0\rangle \;.
  \label{jzerozero} 
\end{equation}
In the rung basis $|\phi_0\rangle$ represents the singlet state, while 
in the plaquette basis it is the 
$|\ell_{14}\!=\!\ell_{23}\!=\!1,j\!=\!0\rangle$ 
single-plaquette wavefunction. The situation becomes slightly more 
complicated for the ($J\!=\!1$) one-magnon states. At zeroth order 
this state is $N$-fold degenerate (with $N\!=\!N_R$ or $N\!=\!N_P$) 
so Eq.~\ref{pertn} is no longer applicable. However, if 
we assume periodic boundary conditions, the total linear momentum 
becomes a good quantum number and serves to distinguish among the 
$N$ degenerate states, which we now write as:
\begin{equation}
  |\Phi_{J=1}^{(\nu)}\rangle = {1\over \sqrt{N}}\sum_{j=1}^{N} 
  \exp \left[i{{2\nu\pi}\over{N}}j \right]|\Phi_{J=1}^{(j)}\rangle \;,
 \label{jonezero} 
\end{equation}
where the linear-momentum label runs from $\nu=0,\ldots,N-1$ and 
\begin{equation}
  |\Phi_{J=1}^{(j)}\rangle =
  |\phi_0\rangle \otimes |\phi_0\rangle\ldots
  |\phi_1\rangle \ldots|\phi_0\rangle \;.
  \label{jzeroone} 
\end{equation}
Now $|\phi_1\rangle$ is the triplet state for the $j$th rung, or
the $|\ell_{14}\!=\!\ell_{23}\!=\!=\!j\!=\!1\rangle$ single-plaquette 
wave function for the $j$th plaquette. In this way, Eq.~\ref{pertn} 
can be used for each individual state in the degenerate band, as states 
with different linear momenta cannot mix at any order in perturbation 
theory. We have systematically determined that, for sufficiently small 
values of $\lambda$, our perturbative calculations agree essentially 
perfectly at large order with exact results obtained by direct 
diagonalization or Lanczos techniques.

\section{Perturbation Theory in Coupled Bases: Bulk Systems}
\label{sec:bulk}

 The essential concept that we use to extract perturbative quantities 
in the bulk limit from calculations for finite systems is that of 
``connectedness''. It is well known in many-body theory (see, {\it e.g.}, 
Refs.~\cite{gelf90,dandf,fandw,nando}) that any given perturbative 
contribution to the coefficients ($E_{J=0}^{(n)}$) appearing in the 
expansion for the ground-state energy [see 
Eqs.~(\ref{pertenergy},\ref{pertn})] 
consists only of spatially connected terms. This is also the basis of 
the connected-graph expansions alluded to in the Introduction. As is 
well-known~\cite{dandf,fandw,nando}, both connected and disconnected 
terms appear in Rayleigh-Schr\"odinger perturbation theory. However,
the disconnected terms are canceled by contributions included in the 
subtractions appearing in the evaluation of the $A_{\alpha\beta}^{(n)}$, 
as shown in Eq.~(\ref{pertn}). The linked-cluster expansions were, of 
course, developed to avoid dealing with disconnected terms which would
eventually be canceled anyway, the price being much greater
conceptual complexity.

In this work, we employ Rayleigh-Schr\"odinger perturbation theory and
find that the computational penalty for not excluding disconnected
contributions from the outset is insignificant compared to the
advantages arising form the simplicity of the formulation. Still, 
we must exploit the connectedness of the surviving contributions to
extract bulk properties. In the present context, connected terms are
those for which a single, finite cluster of adjacent plaquettes can be
identified such that: 1) all interactions involve plaquettes within
that cluster and 2) all plaquettes in the cluster are involved in one
or more interactions (note that in the present discussion we make
explicit reference to plaquettes; however, all the arguments remain
valid in the rung basis, unless stated otherwise). It is apparent
that, at order $n$ in perturbation theory, the {\it largest} such 
cluster contains $n+1$ plaquettes. Hence, we may conclude that, at 
$n$th order, no new kinds of clusters appear as we change the number 
of plaquettes, $N$, provided $N\geq n+1$. Instead, only the number 
of times that each type of cluster appears changes with $N$. As the 
energy must scale at most with $N$, we may conclude that the 
ground-state energy is generally expressible as:
\begin{equation}
  E_{J=0}^{(n)}(N)=A_{0} + B_{0} N \;,
  \label{escale} 
\end{equation}
where $A_{0}$ and $B_{0}$ are ``intensive'' coefficients, which are 
independent of $N$, for $N\geq n+1$. This simple, physically
reasonable behavior---and more specifically the absence of terms 
depending on $N^2$, $N^3$, $\ldots$, is again related to 
connectedness~\cite{dandf,nando}. Indeed, it was in order to eliminate 
terms with ``anomalous'' $N$-dependence that the linked-cluster 
expansion was first proposed.
 
It is clear that, for the case of periodic boundary conditions, we may
set $A_0=0$. Thus, to $n$th order in perturbation theory, the 
ground-state energy-per-site of an $N$-plaquette ladder becomes 
independent of the number of plaquettes for $N \geq n+1$. Obviously 
then, the $n$th order energy-per-site 
{\it in the bulk limit} can be expressed as:
\begin{equation}
  E_{J=0}^{(n)}/{\rm site} = {1 \over 4}
                             E_{J=0}^{(n)}(N\!=\!n\!+\!1)/(n+1) \;,
  \label{egsn} 
\end{equation}
where $E_{J=0}^{(n)}(N)$ is the singlet ground state for an
$N$-plaquette ladder, assuming periodic boundary conditions. 

The structure of the unperturbed one-magnon states 
$(|\Phi_{J=1}^{(\nu)}\rangle)$ or, equivalently, the fact that the
gap is an intensive quantity, ensures that, in the bulk limit, 
the energy-per-site equals that of the ground state. Hence, we may 
conclude that all one-magnon states have the same {\it extensive} 
contribution as the ground state. Therefore,
\begin{equation}
  E_{J=1}^{(n)}(\nu,N)=A^{(\nu)}_{1} + B_{0} N \;.
  \label{eonescale} 
\end{equation}
Again, to $n$th order, $A^{(\nu)}_{1}$ can be determined from studying 
a finite ladder with at least $N\!=\!n\!+\!1$ plaquettes. For $\nu=0$, 
$A^{(\nu\!=\!0)}_{1}$ is just the spin gap at order $n$ and its value 
{\it in the bulk limit} is found simply by taking the difference 
between the $n$th order energies for the lowest one-magnon state and 
the ground state where both are evaluated for an $N=(n+1)$-plaquette 
system with periodic boundary conditions:
\begin{equation}
  \Delta^{(n)}=E_{J=1}^{(n)}(\nu\!=\!0,N\!=\!n\!+\!1) -
               E_{J=0}^{(n)}(N\!=\!n\!+\!1) \;. 
 \label{gapn} 
\end{equation}
We use this formula to determine the perturbative contributions to the 
spin gap in the bulk limit presented in the following sections. Note 
that in the rung basis, it is the linear momentum corresponding to 
$\nu\!=\!N/2$ (i.e., $k=\pi$) which generates the lowest one-magnon 
state. In this case one must employ a finite ladder containing an even 
number of rungs.

It is also easy to show that the $n$th order one-magnon dispersion 
relation has the following form: 
\begin{equation}
  \Delta^{(n)}(k)=\sum_{x=0}^n\,\Delta^{(n)}_x \cos(kx) \;,
 \label{dispn} 
\end{equation}
where $0\leq k \leq 2\pi$ is the linear momentum in lattice units. 
Note that that in the plaquette basis the gap is given by
$\Delta^{(n)}=\Delta^{(n)}(k=0)$ while, in the rung basis, 
$\Delta^{(n)}=\Delta^{(n)}(k=\pi)$. It might be anticipated that, 
at order $n$, the $n+1$ coefficients which specify the full dispersion 
relation at this order could be extracted directly from the values of 
the differences  
\begin{equation}
  \Delta^{(n)}(k=2\nu\pi/N)=E_{J=1}^{(n)}(\nu,N\!=\!n\!+\!1) 
                           -E_{J=0}^{(n)}(N\!=\!n\!+\!1) \;. 
  \label{dispna} 
\end{equation}
However, a moment's thought reveals that some of these one-magnon states 
are degenerate and, thus, energies for systems larger than $N\!=\!n+1$ 
are typically required. Indeed, for a $2 \times 8$ (eight-rung) system, 
a perturbative calculation in the rung basis determines the bulk values 
of the energy-per-site and the gap up to 7th-order in perturbation theory,
but only to 4th-order for the dispersion relation. Nevertheless, it is 
always possible to ascertain the {\it width} of the one-magnon band at 
$n$-th order from calculations for systems with $N\!=\!n\!+\!1$ 
plaquettes, provided $N$ is even.

Finally, it should be noted that the structure of the ground state in
the plaquette basis (and, indeed, in any basis in which the
unperturbed ground state is made out of singlets) affords a less
obvious but more efficient means of extracting the coefficients that
determine the ground-state energy in the bulk limit. To see this, we
first note that $|\Phi_{J=0}\rangle$ contains only $j\!=\!0$
plaquettes [see Eq.~(\ref{jzerozero})]. Since the inter-plaquette
interaction $V$, specified in Eq.~(\ref{plaquettevee}), consists of
rank-1 operators in the space of a single plaquette, any plaquette
appearing in a connected cluster must experience at least {\it two}
interactions in order that it be returned to its original $j\!=\!0$
state. For this reason, it can be shown that $E_{J=0}^{(n)}(N)$, when
computed with {\it open} boundary conditions, may be expressed as:
\begin{equation}
  E_{J=0}^{(n)}(N)=\sum_{m=1}^{[n/2]} 
                  (N-m)\,W_m^{(n)}(N\!=\!m\!+\!1) \;,
  \label{egsopen} 
\end{equation}
where $[n/2]$ is the largest integer less than or equal to $n/2$. 
Then, for example, the quantities $W_{1}^{(5)}$ and $W_{2}^{(5)}$ can be 
extracted from fifth-order calculations for systems with only two and three 
plaquettes. (Note that, in contrast, the extraction of fifth-order 
coefficients ($E_{J=0}^{(n)}$) for the ground-state energy using 
periodic-boundary conditions and Eq.~(\ref{egsn}) requires calculations 
for systems with six plaquettes.) Then, the $n$th-order energy-per-site in 
the bulk limit is easily found to be
\begin{equation}
  E_{J=0}^{(n)}/{\rm site} = {1 \over 4} 
  \sum_{m=1}^{[n/2]} W_m^{(n)}(N\!=\!m\!+\!1) \;.
 \label{epsbulk} 
\end{equation}
We use this formula to determine the bulk perturbative ground state 
energy contributions presented in the following sections.

\section{Results}
\label{sec:results}

\subsection{The Rung Basis}
\label{sec:resrungbasis}

\begin{figure}
\centerline{
 \psfig{figure=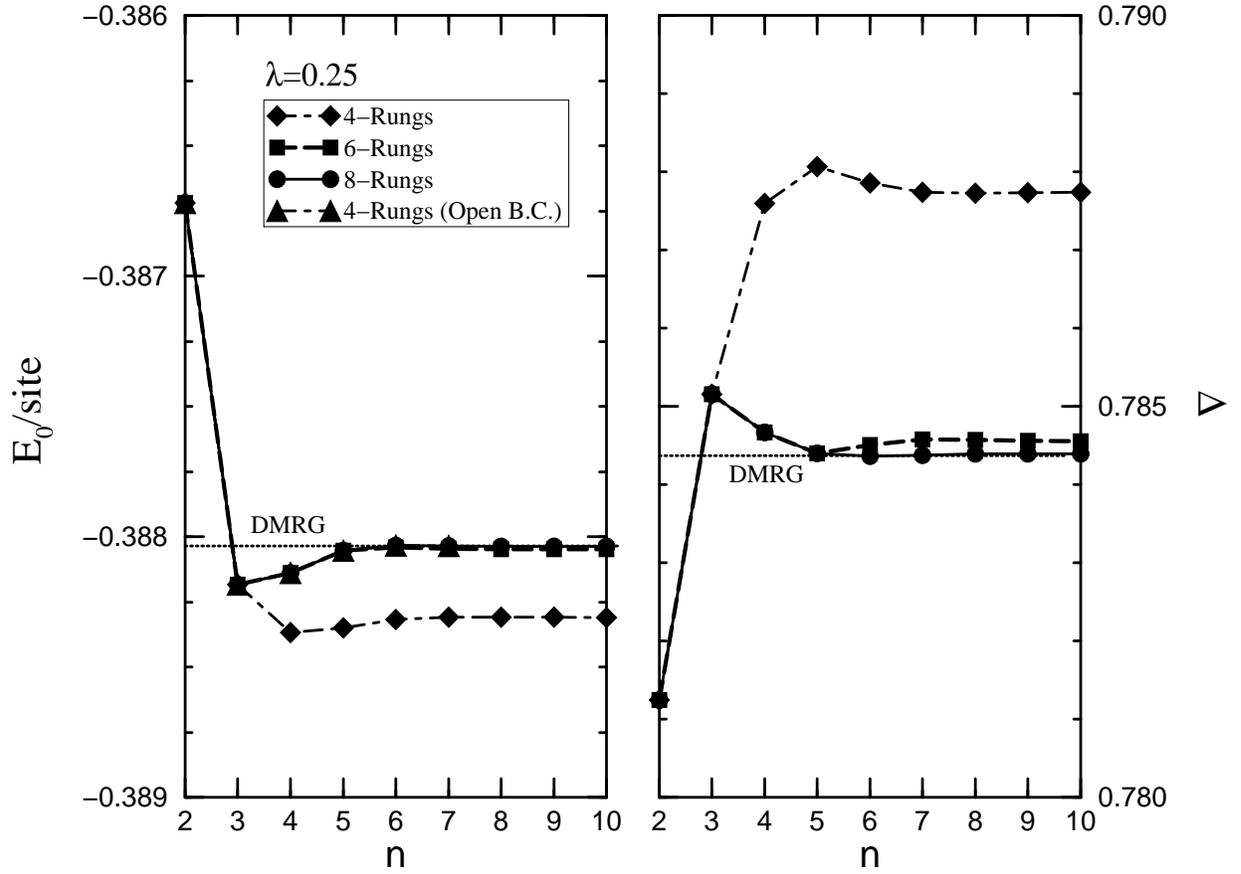,height=5.5in,width=6.5in,angle=-90}}
\caption{Rung-basis perturbation theory results through order $n=10$ 
	 for the ground-state 
         energy per site (left) and for the spin gap (right) for systems 
         with 4, 6 and 8 rungs using periodic boundary conditions. For
         the ground-state energy, the triangles display a 4-rung
	 calculation with open boundary conditions (see text for an
         explanation). Here $\lambda=0.25$ and the DMRG  values are from 
	 Ref.~\protect\cite{lkamp98}}. 
\label{figonerung}
\end{figure}

We first demonstrate our method for extracting perturbative energies
in the bulk limit from Rayleigh-Schr\"odinger calculations for finite
systems. Fig.~\ref{figonerung} shows rung-basis calculations for the 
ground state energy and for the spin gap for systems with 4, 6, and 8 
rungs. These calculations were done for $\lambda=0.25$ so convergence 
to DMRG values\cite{lkamp98} is very fast. Fig.~\ref{figonerung} shows
that, at orders $n=2$ and $3$, all calculations give the same per-rung
energy and spin-gap contributions, as we would expect from the arguments 
of the preceding section. At $n=4$, the 6- and 8-rung values coincide, 
but now differ from the 4-rung results, again consistent with our 
expectations. Finally, the 6- and 8-rung values begin to differ for 
$n\geq 6$. Note that our 7th-order perturbative result for the 
spin gap agrees with the DMRG value to about 0.07\% 
(for the energy per site both results are indistinguishable
to five significant figures). The left-hand side of 
Fig.~\ref{figonerung} also demonstrates that, when Eq.~(\ref{epsbulk})
is used, the contributions to the ground-state energy extracted from 
4-rung calculations with open boundary conditions agree with those 
obtained from an 8-rung ladder with periodic boundary conditions 
through $n\!=\!7$, as expected. This result embodies the power of the 
method, and the high degree of symmetry of the unperturbed ground
state and the residual interaction. Indeed, an 8-rung calculation 
affords the computation of the ground-state energy---up to 15th order 
in perturbation theory.

Moreover, from one such calculation, namely, a $2\times 8$ ladder at
a single value of $\lambda$, one can extract the perturbative 
coefficients for the ground-state energy and the spin gap up to 
7th-order. That is, the ground-state energy per site is given by
\begin{equation}
    E_{0}/N = -{  3 \over 8}
              -{  3 \over 16}  \lambda^2
              -{  3 \over 32}  \lambda^3
              +{  3 \over 256} \lambda^4
              +{ 45 \over 512} \lambda^5
              +{159 \over 2048}\lambda^6
              -{ 55 \over 2048}\lambda^7 \;,
 \label{eseventh}
\end{equation}
while for the spin gap we obtain:
\begin{equation}
     \Delta =     1   -        \lambda
              +{  1 \over    2}\lambda^2
              +{  1 \over    4}\lambda^3
              -{  1 \over    8}\lambda^4
              -{ 35 \over  128}\lambda^5
              -{157 \over 1024}\lambda^6
              +{503 \over 2048}\lambda^7 \;.
 \label{gseventh}
\end{equation}

\begin{figure}
\centerline{
 \psfig{figure=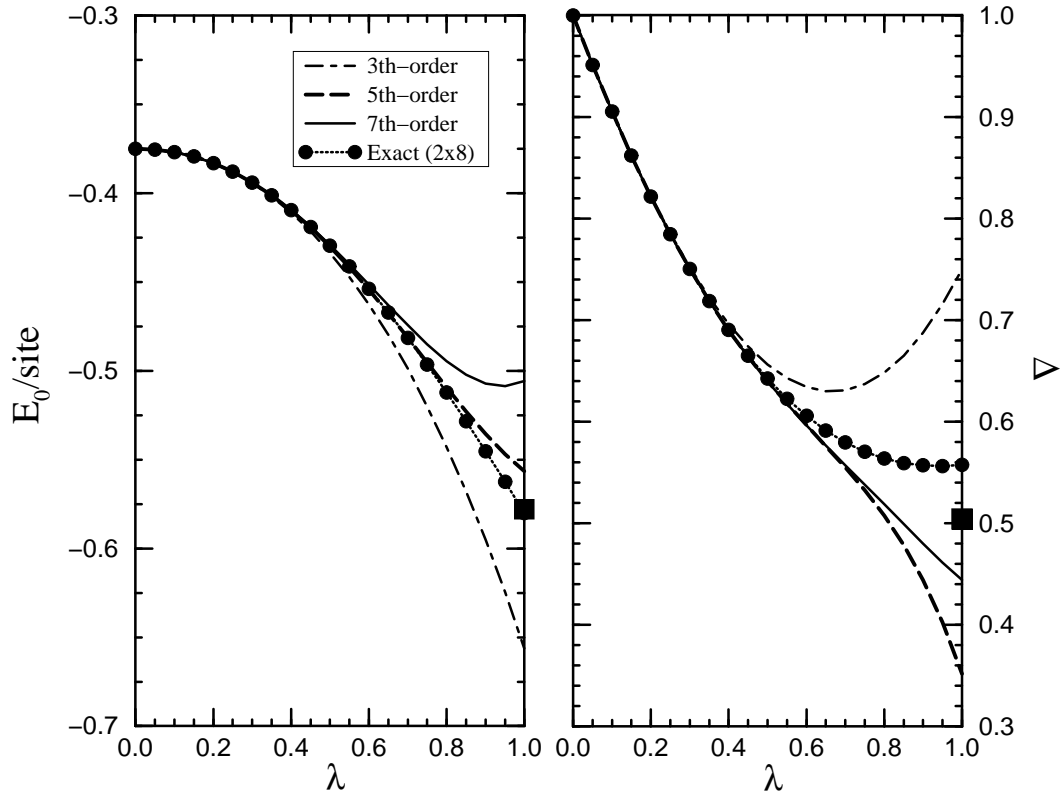,height=5in,width=6in,angle=-90}}
\caption{Ground-state energy per site (left) and spin gap (right) in
	 the bulk limit as a function of $\lambda$ in the rung basis.
         Various orders of perturbation theory are displayed as well
         as the exact results on an 8-rung ladder. The DMRG 
	 values\protect\cite{white94} for the bulk system at the 
	 isotropic point ($\lambda=1$) are displayed with a solid square.}
\label{figtworung}
\end{figure}

Of course, these two expressions can be used for arbitrary values of
$\lambda$, provided these values lie within the radius of convergence
of the perturbative expansions (more on this point later). In
Fig.~\ref{figtworung} we have displayed the bulk values for the
ground-state energy per site and the spin gap to various orders in
perturbation theory; we have also included exact results obtained on a
$2\times 8$ ladder. We observe that in the strong-coupling regime
$(\lambda\alt 0.5)$ the convergence of our results is very good. This
should not come as a surprise, as the rung basis has been customized
to handle the strong-coupling region. In contrast, the convergence of
the perturbative results in the intermediate- to isotropic-coupling region
$(\lambda\agt 0.8)$ is harder to assess. However, the fact that there is
no systematic improvement---nor convergence---in the results as one
goes to higher orders in perturbation theory, strongly suggests that the
expansion might no longer be convergent.

\begin{figure}
\centerline{
 \psfig{figure=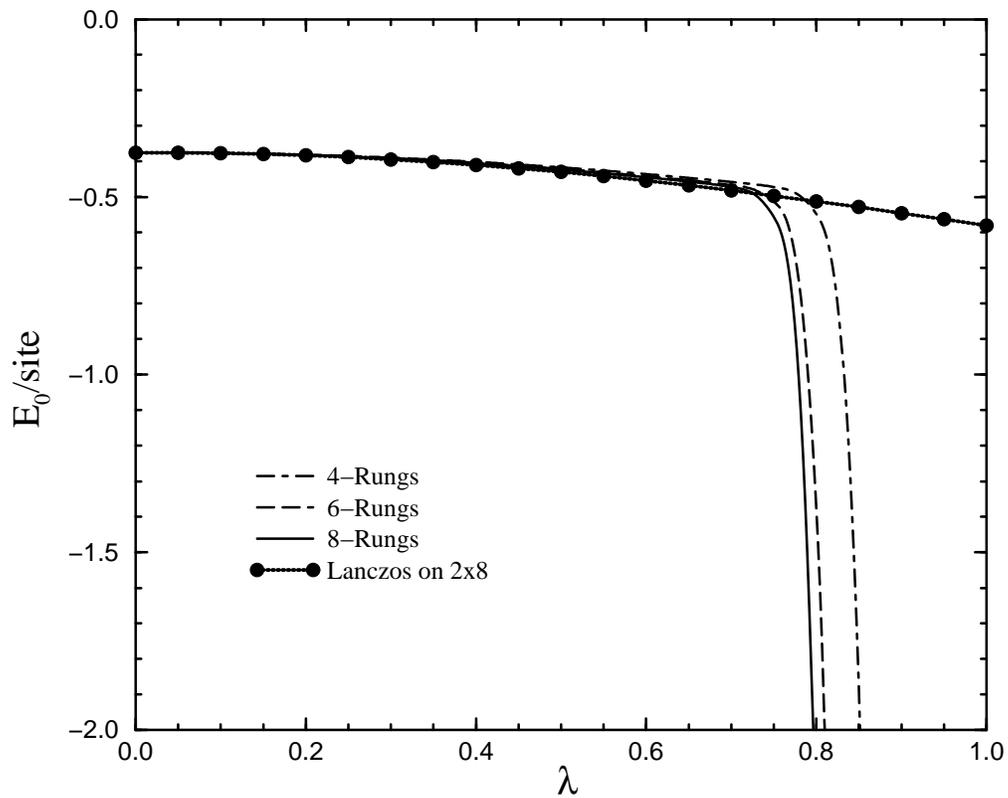,height=5in,width=6in,angle=-90}}
\caption{Ground state energy per site for 4-, 6-, and 8-rung ladders
	 at 50th order in perturbation theory as a function of 
	 $\lambda$. We conclude that the radius of convergence 
         of the rung-basis perturbative expansion is about 
	 $\lambda_c=0.8$.}

\label{figthreerung}
\end{figure}

In an effort to establish quantitatively the radius of convergence of 
the perturbative expansion, we have compared high-order perturbative 
and exact results for finite ladders over a range of $\lambda$ values. 
For example, Fig.~\ref{figthreerung} shows the ground-state energy per
site through 50th order in perturbation theory for 4-, 6- and 
8-rung ladders as a function of $\lambda$. The radius of 
convergence is determined by the onset of wild excursions in these 
quantities at $\lambda_c\simeq 0.8$. These perturbative results are 
indistinguishable from the Lanczos energies for $\lambda\leq \lambda_c$. 
As the value of $\lambda_c$ appears to have very little dependence on 
the size of the ladder, we assume it is also appropriate to the bulk 
system. Thus, we conclude that the radius of convergence of the 
perturbative expansion in the rung basis is $\lambda_c=0.8$. This
result indicates that perturbative expansions in the rung basis  
are unsuitable for dealing with the physically-interesting case of
isotropic $(\lambda=1)$ ladders.

 Finally, we conclude this section by listing in Table~\ref{tableonerung}
the perturbative coefficients for the one-magnon dispersion relation, as
given in Eq.~(\ref{dispn}).

\subsection{The Plaquette Basis}
\label{sec:resplaqbasis}

\begin{figure}
\centerline{
  \psfig{figure=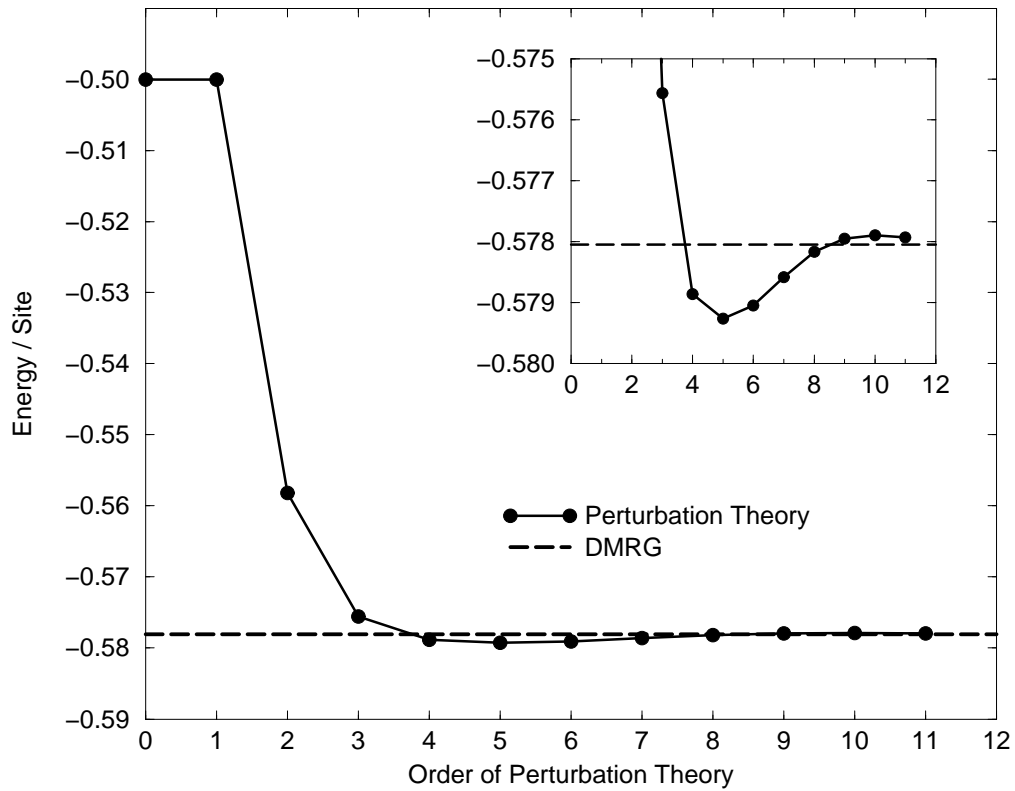,height=5in,width=6in,angle=-90}}
\caption{ Ground state energy per site in the bulk limit versus order  
	  of perturbation theory in the plaquette basis. The inset shows
	  the same results but on an expanded scale. The DMRG value is
	  from Ref.~\protect\cite{white94}.}
\label{figoneplaq}
\end{figure}

\begin{figure}
\centerline{
 \psfig{figure=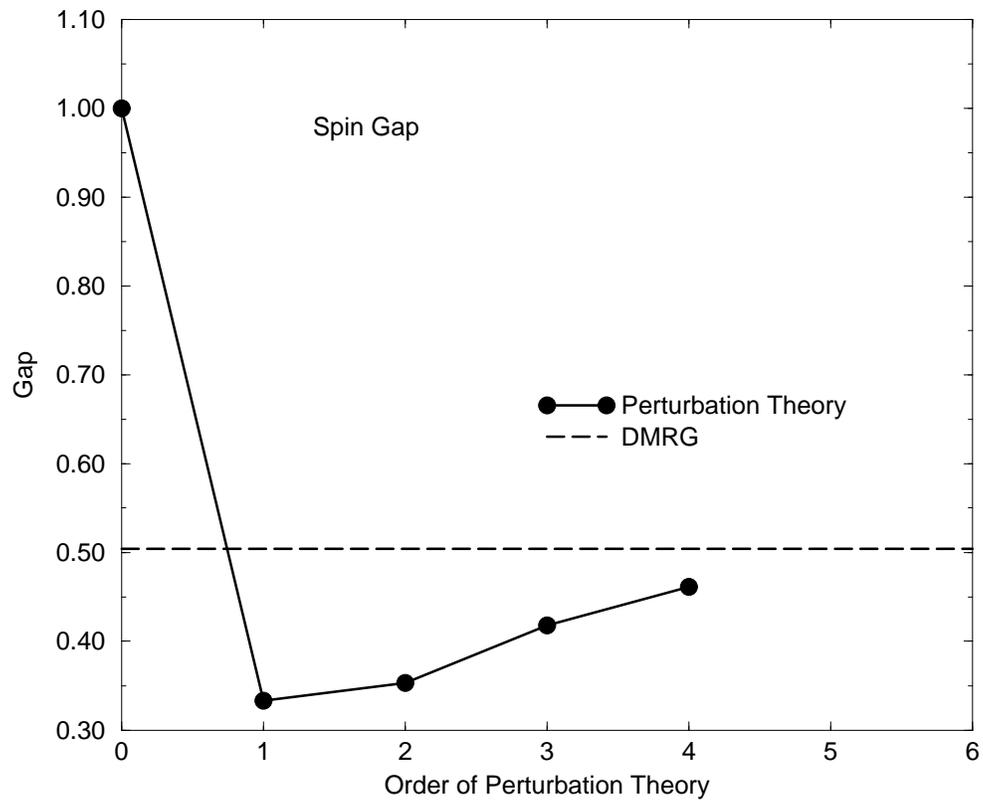,height=5in,width=6in,angle=-90}}
\caption{ Same as Fig.~\ref{figoneplaq} but for the spin gap.}
\label{figtwoplaq}
\end{figure}

 Plaquette-based perturbative results for the ground-state energy
per site and for the spin gap appear in Table~\ref{tableoneplaq}. 
By carrying out Rayleigh-Schr\"odinger calculations for isotropic
$(J_\parallel=J_\perp)$ ladders as  large as $2 \times 12$ 
({\it i.e.}, six plaquettes), we have been able to determine the 
ground-state energy per site through 11th order and the spin gap 
through fourth order. Through 11th order, the ground-state energy
is within 0.02\% of the DMRG~\cite{white94} value while at fourth 
order, the perturbative estimate for the spin gap agrees with the 
DMRG value to better than 10\%. The perturbative energy 
per site is plotted versus the order of perturbation in 
Fig.~\ref{figoneplaq} where the rapid convergence to the DMRG value 
is apparent. A similar plot for the spin gap is shown in 
Fig.~\ref{figtwoplaq}. Here convergence is less dramatic presumably 
due to the relatively low order of perturbation theory to which 
we are restricted. As became apparent in the previous section, 
there is no point in presenting a similar table or figures for 
rung-basis perturbation theory, as it does not converge for 
$\lambda=1$ which is the only point where direct rung versus 
plaquette comparisons are possible. 

\begin{figure}
\centerline{
 \psfig{figure=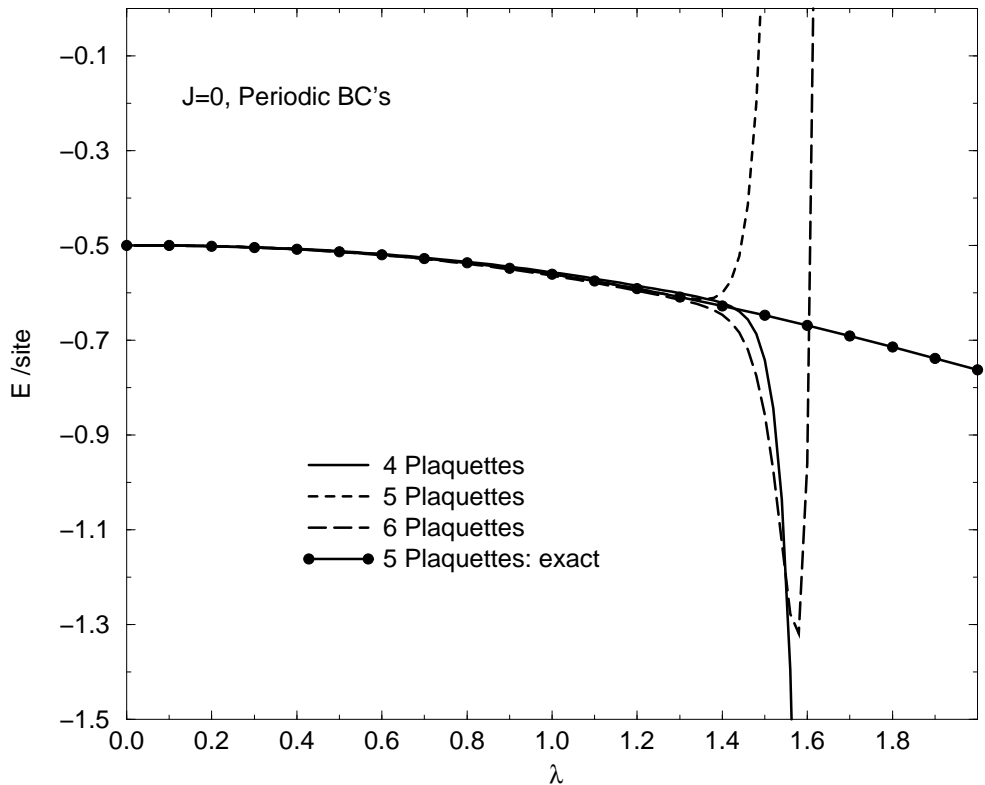,height=5in,width=6in,angle=-90}}
\caption{ Ground state energy per site for 4-, 5-, and 6-plaquette systems 
	  at 50th order in perturbation theory as a function of $\lambda$
	  which scales the inter-plaquette interaction (see the subsection on 
	  Rayleigh-Schr\"odinger Perturbation Theory).
          We conclude that the radius of convergence of the plaquette-basis 
          perturbative expansion is about $\lambda_c=1.4$.}
\label{figthreeplaq}
\end{figure}

\begin{figure}
\centerline{
\psfig{figure=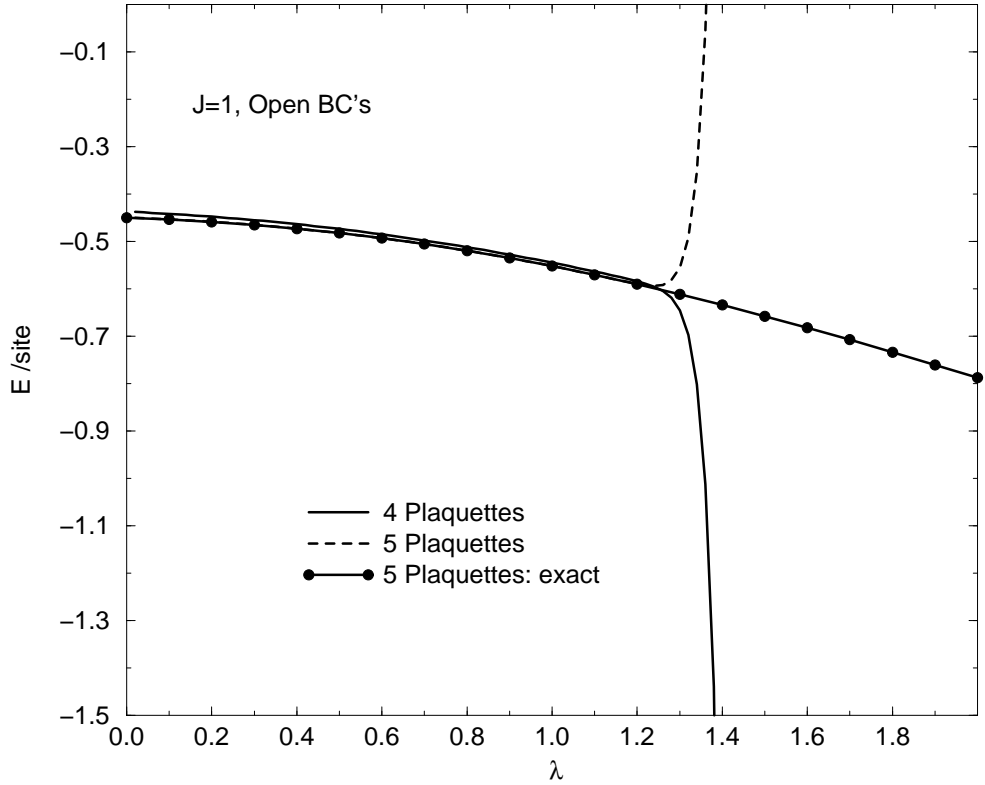,height=5in,width=6in,angle=-90}}
\caption{ Same as Fig.~\ref{figthreeplaq} but for the lowest triplet state. 
	  We conclude that the radius of convergence of the plaquette-basis 
	  perturbative expansion here is about $\lambda_c=1.25$.}
\label{figfourplaq}
\end{figure}

 To extract the radius of convergence in plaquette-based perturbation 
theory we have again compared high-order perturbative and exact
results for finite ladders over a range of $\lambda$ values. Indeed,
Fig.~\ref{figthreeplaq} shows the ground-state energy per site through
50th order in perturbation theory for 4-, 5- and 6-plaquette systems 
as a function of $\lambda$. Notice that for the present case the onset
of the wild excursions is now at $\lambda_c\simeq 1.4$; once more the 
agreement with exact calculations using direct diagonalization or 
Lanczos methods is essentially perfect for $\lambda\leq \lambda_c$.
Fig.~\ref{figfourplaq} shows similar plots for the lowest triplet 
state of ladders with 4 and 5 plaquettes. On the basis of the behavior
shown, we estimate that $\lambda_c\simeq 1.25$ for this state. It is 
therefore straightforward to conclude that the plaquette basis 
offers, to date, the only reliable perturbative approach for the 
treatment of the physically interesting\cite{dagric96} case of 
isotropic ($\lambda=1$) spin ladders.

\begin{figure}
\centerline{
 \psfig{figure=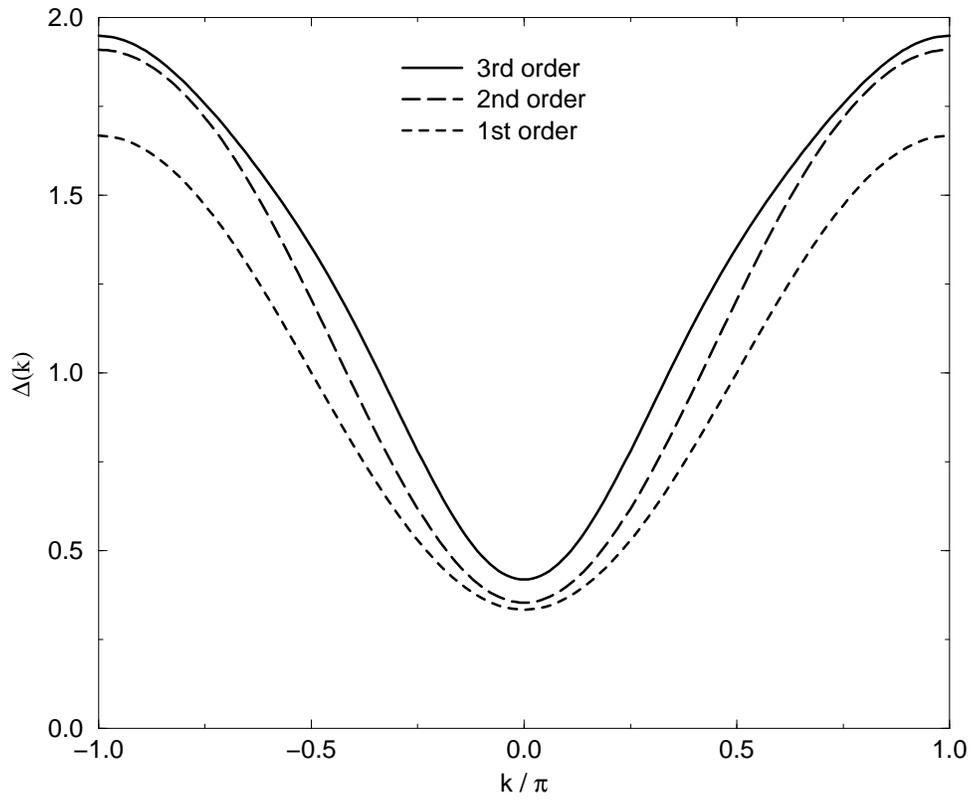,height=5in,width=6in,angle=-90}}
 \caption{One-magnon dispersion relation at various orders of 
	  plaquette-basis perturbation theory.}
 \label{figfiveplaq}
\end{figure}

 Finally we conclude this section by presenting in
Fig.~\ref{figfiveplaq} the perturbative one-magnon dispersion relation
for the plaquette basis. The curves presented in these figures were 
generated by inserting the coefficients appearing in 
Table~\ref{tabletwoplaq} into Eq.~(\ref{dispn}). 

\section{Summary and Conclusions}
\label{sec:concl}
 
We have employed rung and plaquette bases to study bulk properties of
Heisenberg spin-1/2 ladders using Rayleigh-Schr\"odinger perturbation
theory. In both cases, the mere selection of an angular momentum coupled 
basis enables one to separate
the Hamiltonian into a unperturbed part, which is solved exactly, and 
a ``residual interaction'' which is treated in perturbation theory. In
this way the starting point for the calculation are unperturbed 
wavefunctions consisting of a direct product of spin singlets for
the ground state, and a degenerate ``one-magnon'' band in which one
spin singlet has been turned into a triplet. The unperturbed 
ground-state energies per site in the rung and plaquette bases are 
given by $-3/8$ and $-1/2$, respectively; the spin gap starts at 
$\Delta=1$ in both cases.

The one-dimensional nature of the problem enabled us to employ
powerful methods for the extraction of high order perturbative
coefficients for the bulk system from calculations on relatively small 
ladders. Indeed, using very general arguments---related to the 
linked-cluster expansion and to the scaling of observables in the bulk 
limit---we have shown that from a calculation on a ladder with $N$ 
clusters (a cluster being a rung or a plaquette) we can extract all 
perturbative coefficients for the ground-state energy and for the spin 
gap up to order $N-1$. Moreover, with a simple modification, namely using 
open-boundary conditions, the ground-state energy per site could be 
computed up to order $2N-1$. Yet, because of degeneracies in 
the one-magnon band, perturbative coefficients for the one-magnon 
dispersion relation could only be extracted up to order $N-2$ by 
combining results from Rayleigh-Schr\"odinger calculations for ladders
with $N$ and $N-1$ plaquettes.

For the rung basis we have carried out calculations on a relatively
small $2\times 8$ ladder; this sort of computations can now be carried
out routinely on a personal computer. Yet, the power of the method 
allowed us to evaluate the ground-state energy and the spin gap to 
7th-order in perturbation theory. Perhaps as important, we could 
reliably estimate the radius of convergence of the perturbative 
expansion in the rung basis. This is essential, as such expansions 
are routinely assumed to be meaningful in the isotropic 
($J_\parallel/J_\perp\sim 1$) limit~\cite{barnes93,reig94,gopa94}. 
We have found a radius of convergence for the perturbative expansion 
in the rung basis of only $\lambda=J_\parallel/J_\perp\simeq 0.8$.

Similar calculations were also carried out in the plaquette basis.
Here the parameter $\lambda$ scales the inter-plaquette to
the intra-plaquette interactions; the isotropic
$(J_\parallel=J_\perp)$ limit was obtained by setting $\lambda=1$.
We were able to perform calculations for the singlet ground state 
on a $2\times 12$ ladder. For the triplet states, we were restricted to a 
$2\times 10$ ladder.
Even when a ladder of such large size was employed, the spin gap could 
only be computed up to 4th-order in perturbation theory; the ground-state
energy was computed with open-boundary conditions so 11th-order
results were reported. At first sight this appears as a drawback
relative to the rung-basis calculations described earlier. However, on
close examination one observes that rung-basis perturbation expansions
in the physically-interesting case of the isotropic limit
$(\lambda=1)$ are not feasible even in principle.  In contrast, the
isotropic point is well within the radius of convergence---estimated
to be at $\lambda_c\simeq 1.25$---for plaquette-based
expansions. Finally, by a combination of 4- and 5-plaquette results,
we were able to extract the one-magnon band dispersion relation through 
3rd order in perturbation theory.

In conclusion, we have formulated perturbation theory for two-leg 
Heisenberg spin ladders in two angular momentum coupled bases, namely
the ``rung'' and ``plaquette'' bases. By exploiting the connectedness
of perturbative energy contributions, we have been able to 
straightforwardly extract quantities appropriate to the bulk system from 
Rayleigh-Schr\"odinger perturbation theory for finite ladders. We find that
the radius of convergence in the rung basis is $\lambda_c\simeq 0.8$ while,
in the plaquette basis, it is $\lambda_c\simeq 1.25$. Clearly, the latter 
basis is strongly preferred when treating the physically interesting
case of {\it isotropic} spin ladders which corresponds to $\lambda_c = 1$ 
in both bases. As we have stressed elsewhere\cite{piek96,piek97}, the 
plaquette basis is especially well-suited to the study of many aspects 
of spin ladders. Its success in the context of perturbation theory as 
presented here encourages us to use it in other investigations 
including a mean-field treatment
of ladders (see, {\it e.g.}, Ref.~\cite{gopa94} for a rung-basis approach)
and in a $t-J$ model aimed at the very interesting question of the dynamics 
of doped ladders (again, see, {\it e.g.}, Ref.~\cite{sig94} for a rung-basis 
approach). These studies will be the subject of future reports.

\acknowledgments
We thank Martin Gelfand for useful discussions and for providing us
with his unpublished plaquette results obtained independently using
different techniques. We also thank Markus Laukamp for making
available the DMRG results. This work was supported by the DOE under
Contracts Nos. DE-FC05-85ER250000, DE-FG05-92ER40750 and
DE-FG03-93ER40774.


\vfill\eject
\vfill\eject
\vfill\eject
\vfill\eject
\vfill\eject
\vfill\eject
\vfill\eject
\vfill\eject
\vfill\eject

\mediumtext
 \begin{table}
  \caption{Rung-based perturbation theory for the one-magnon dispersion 
	   relation; see Eq.~(\ref{dispn}).}
   \begin{tabular}{cccccc}
   Order $n$ & $\Delta^{(n)}_0$ & $\Delta^{(n)}_1$ & $\Delta^{(n)}_2$ 
             & $\Delta^{(n)}_3$ & $\Delta^{(n)}_4$ \\
     \tableline
 0 &   1           &                &             &              &   \\
 1 &   0           &  1             &             &              &   \\
 2 &   3/4         &  0             &  $  -1/4$   &              &   \\
 3 &   3/8         & $-1/4 $        &  $  -1/4$   &  1/8         &   \\
 4 & $-13/64$      & $-5/16$        &  $-5/160$   &  1/8   
   & $ -5/64$      \\
   \end{tabular}
  \label{tableonerung}
 \end{table}
\mediumtext
 \begin{table}
  \caption{Plaquette-based perturbation theory for the bulk energy per
           site and spin gap of isotropic 2-leg spin ladders. 
           [see Eq.~(\ref{pertenergy})] DMRG results are from 
	   Ref.~\protect\cite{white94}.}
   \begin{tabular}{ccc}
     Order $n$  &  $E^{(n)}_0$ & $\Delta^{(n)}$ \\
     \tableline
        0     &  -0.5        &    1           \\
        1     &     0        &   -2/3         \\
        2     &  -0.05815972 &    0.01967593  \\
        3     &  -0.01739728 &    0.06493538  \\
        4     &  -0.00329927 &    0.04306155  \\
        5     &  -0.00040692 &                \\
        6     &   0.00021569 &                \\
        7     &   0.00046854 &                \\
        8     &   0.00041483 &                \\
        9     &   0.00021860 &                \\
       10     &   0.00005215 &                \\
       11     &  -0.00003610 &                \\
     \tableline
      Total    &  -0.57792948 &    0.46100619  \\
     \tableline
     DMRG     &  -0.57804314 &    0.504       \\
   \end{tabular}
  \label{tableoneplaq}
 \end{table}

\mediumtext
 \begin{table}
  \caption{Plaquette-based perturbation theory for one-magnon dispersion 
	   relation; see Eq.~(\ref{dispn}).}
   \begin{tabular}{ccccc}
   Order $n$ & $\Delta^{(n)}_0$ & $\Delta^{(n)}_1$ & $\Delta^{(n)}_2$ 
             & $\Delta^{(n)}_3$ \\
     \tableline
 0 & 1           &                &             &                        \\
 1 & 0           &  -2/3          &             &                        \\
 2 & 145/864     &  -1/9          & -1/27       &                        \\
 3 & 0.10022425  &  0.04997348    & -0.04822531 & -1/27                  \\
   \end{tabular}
  \label{tabletwoplaq}
 \end{table}

\end{document}